\documentstyle [12pt]{article}

\def\fnote#1#2{\begingroup\def\thefootnote{#1}\footnote{#2}\addtocounter
{footnote}{-1}\endgroup}
\def\txs{\textstyle} \def\half{{\txs{1\over2}}}
\def\np#1{{\sl Nucl.~Phys.~\bf B#1}}\def\pl#1{{\sl Phys.~Lett.~\bf #1B}}
\def\pr#1{{\sl Phys.~Rev.~\bf D#1}}\def\prl#1{{\sl Phys.~Rev. Lett.~\bf #1}}
\def\cpc#1{{\sl Comp.~Phys. Comm.~\bf #1}} \def\be{\begin{equation}}
\def\anp#1{{\sl Ann.~Phys.~(NY) \bf #1}} \def\ee{\end{equation}\ }
\def\etal{{\it et al.}}\def\sqstev{$\sqrt s=40$~TeV}
\def\bbar#1{\llap{\phantom#1}^{\scriptscriptstyle(}\bar#1^{\scriptscriptstyle)}}
\def\Kmax{K_{\rm max}}\def\ieps{{i\epsilon}}\def\rQCD{{\rm QCD}}
\def\rngp{renormalization group\ }
\begin{document}
\hfill UTHEP--93--0401\vskip.01truein
\hfill {April 1993}\vskip1truein
\centerline{\Large Renormalization Group Improved Exponentiation}
\centerline{\Large of Soft Gluons in QCD\fnote{\ast}{Research supported in part
by the Texas National Research Laboratory Commission grants RCFY9101, RCFY9201,
and RCFY-93-347 and by Polish Government grants KBN 20389101 and KBN
223729102.}}\vskip.8truein
\centerline{\sc D.~B.~DeLaney, S.~Jadach,\fnote{\dagger}{Permanent address:
Institute of Nuclear Physics, ul. Kawiory 26a, Cracow, Poland.} Ch.~Shio,
G.~Siopsis, and B.~F.~L.~Ward}\vskip.5truein
\centerline{\it Department of Physics and Astronomy}
\centerline{\it The University of Tennessee, Knoxville, TN 37996--1200}
\vskip.05truein \baselineskip=21pt\vskip1.1truein
\centerline{\bf ABSTRACT}\vskip.2truein\par
We extend the methods of Yennie, Frautschi and Suura to QCD for the summation
of
soft gluon effects in which infrared singularities are cancelled to all orders
in $\alpha_s$. An explicit formula for the respective \rngp improved
exponentiated cross section is obtained for $q+\bbar{{q'}}\to q+\bbar{{q'}}+
n(G)$ at SSC energies. Possible applications are discussed.
\par\renewcommand\thepage{}\vfill\eject
\parskip.1truein \parindent=20pt \pagenumbering{arabic}
\par
As the SSC is now being constructed, it is an important problem to compute
reliably
the higher order radiative corrections to the relevant physics processes
which it will probe either as signal or as background. Recently~\cite{ref1}, we
have shown that the higher order radiative corrections to SSC processes
generated by multiple photon effects can be handled by the
Yennie-Frautschi-Suura (YFS)~\cite{ref2} methods which two of us~\cite{ref3}
(S.J. and B.F.L.W.) have realized via Monte Carlo methods in the context of
high
precision $Z^0$ physics. The outstanding issue is to extend these YFS Monte
Carlo exponentiation methods to QCD multiple gluon radiative effects. In this
Letter, we prove that such an extension exists. Its realization in an explicit
Monte Carlo event generator will appear elsewhere~\cite{ref4}.\par
Specifically, we want to compute the QCD analogues of the YFS infrared
functions
$\tilde B$ and $B$, which represent the real and virtual infrared singularities
in QED, respectively: at order $\alpha$, the cross sections\be d\sigma^{\rm(1-
loop)}-2\alpha{\rm Re}B\,d\sigma_B\label{one}\ee and \be\int^{k\le\Kmax}dk\,
\bar\beta_1(k)=d\sigma^{B1}-2\alpha\tilde B(\Kmax)\,d\sigma_B\label{two}\ee are
both infrared finite by definition and in addition \be{\rm SUM_{IR}}\equiv2
\alpha{\rm Re}B+2\alpha\tilde B(\Kmax)\label{three}\ee is also infrared finite,
so that the cancellation of infrared singularities is realized to all orders in
$\alpha$ via the exponentiation of (\ref{three}) as a consequence of the
independent emission of arbitrarily soft photons in QED~\cite{ref2}. Here,
$\Kmax$ is the detector resolution soft photon parameter (a photon is soft if
its energy is less than $\Kmax$); $d\sigma^{\rm(1-loop)}$ is the relevant
generic cross section with one-loop virtual corrections; and $d\sigma^{B1}$ is
the corresponding cross section with one real $\gamma$ emission.\par
In QCD, consider our prototypical SSC process $q+\bbar{{q'}}\to q+\bbar{{q'}}+
(G)$ at
order $\alpha_s$. The relevant Feynman graphs and kinematics are shown in
Fig.~1. Following the definitions of $\tilde B$ and $B$ in Ref.~\cite{ref2}, we
get for QCD the results\[B_\rQCD={i\over(8\pi^3)}{\int d^4k\over(k^2-m_G    ^2+
\ieps)}\left[C_F\left({2p_1+k\over k^2+2k\cdot p_1+\ieps}+{2p_2-k\over k^2-2k
\cdot p_2+\ieps}\right)^2\right.\hfil\]\[\hfil+\Delta
C_s{2(2p_1+k)\cdot(2p_2-k)
\over(k^2+2k\cdot p_1+\ieps)(k^2-2k\cdot p_2+\ieps)}\]\be\begin{array}{rl}+C_F
\left({\txs2q_1+k\over\txs k^2+2k\cdot q_1+\ieps}+{\txs 2q_2-k\over\txs k^2-2k
\cdot q_2+\ieps}\right)^2+&\Delta
C_s{\txs2(2q_1+k)\cdot(2q_2-k)\over\txs(k^2+2k
\cdot q_1+\ieps)(k^2-2k\cdot q_2+\ieps)}\\ +C_F\left({\txs2p_2+k\over\txs
k^2+2k
\cdot p_2+\ieps}-{\txs2q_2+k\over\txs k^2+2k\cdot q_2+\ieps}\right)^2-&\Delta
C_t{\txs2(2q_2+k)\cdot(2p_2+k)\over\txs(k^2+2k\cdot q_2+\ieps)(k^2+2k\cdot p_2+
\ieps)}\\ +C_F\left({\txs2p_1+k\over\txs k^2+2k\cdot
p_1+\ieps}-{\txs2q_1+k\over
\txs k^2+2k\cdot q_1+\ieps}\right)^2-&\Delta
C_t{\txs2(2p_1+k)\cdot(2q_1+k)\over
\txs(k^2+2k\cdot p_1+\ieps)(k^2+2k\cdot q_1+\ieps)}\\
-C_F\left({\txs2p_1+k\over
\txs k^2+2k\cdot p_1+\ieps}-{\txs2q_2+k\over\txs k^2+2k\cdot
q_2+\ieps}\right)^2
+&\Delta C_u{\txs2(2p_1+k)\cdot(2q_2+k)\over\txs(k^2+2k\cdot p_1+\ieps)(k^2+2k
\cdot q_2+\ieps)}\\ -C_F\left({\txs2q_1+k\over\txs k^2+2k\cdot
q_1+\ieps}-{\txs2
p_2+k\over\txs k^2+2k\cdot p_2+\ieps}\right)^2+&\Delta C_u\left.{\txs2(2q_1+k)
\cdot(2p_2+k)\over\txs(k^2+2k\cdot q_1+\ieps)(k^2+2k\cdot p_2+\ieps)}\right]
\end{array}\label{four}\ee and \be2\alpha_s\tilde B_\rQCD=\int{d^3k\over k_0}
\tilde S_\rQCD(k)\label{five}\ee with\[\tilde
S_\rQCD(k)=-{\txs\alpha_s\over\txs
4\pi^2}\left\{C_F\left({\txs p_1\over\txs p_1\cdot k}-{\txs q_1\over\txs q_1
\cdot k}\right)^2\right.-\Delta C_t{\txs2p_1\cdot q_1\over\txs k\cdot p_1k\cdot
q_1}\]\[\begin{array}{rccl}+C_F\left({\txs p_2\over\txs p_2\cdot k}-{\txs q_2
\over\txs q_2\cdot k}\right)^2&-\Delta C_t{\txs2p_2\cdot q_2\over\txs k\cdot
p_2
k\cdot q_2}&+C_F\left({\txs p_1\over\txs p_1\cdot k}-{\txs p_2\over\txs
p_2\cdot
k}\right)^2&-\Delta C_s{\txs2p_1\cdot p_2\over\txs k\cdot p_1k\cdot p_2}\\ +C_F
\left({\txs q_1\over\txs q_1\cdot k}-{\txs q_2\over\txs q_2\cdot k}\right)^2&-
\Delta C_s{\txs2q_1\cdot q_2\over\txs k\cdot q_1k\cdot q_2}&-C_F\left({\txs q_1
\over\txs q_1\cdot k}-{\txs p_2\over\txs p_2\cdot k}\right)^2&+\Delta C_u{\txs2
q_1\cdot p_2\over\txs k\cdot q_1k\cdot p_2}\end{array}\]\be-C_F\left({\txs q_2
\over\txs q_2\cdot k}-{\txs p_1\over\txs p_1\cdot k}\right)^2+\left.\Delta C_u{
\txs2q_2\cdot p_1\over\txs k\cdot q_2k\cdot p_1}\right\},\label{six}\ee
\smallskip where $C_F=4/3=$ quadratic Casimir invariant of the quark color
representation, $m_G    $ is a standard infrared regulator mass and
\begin{eqnarray}\Delta C_s&=\left\{\begin{array}{ll}-1\qquad,&qq'\ {\rm
incoming
}\\ -1/6\quad,&q\bar q'\ {\rm incoming}\end{array}\right.&,\quad\Delta
C_t=-3/2,
\ {\rm and}\nonumber\\ \Delta C_u&=\left\{\begin{array}{ll}-5/2\quad,&qq'\ {\rm
incoming}\\ -5/3\quad,&q\bar q'\ {\rm
incoming}\end{array}\right.&,\label{seven}
\end{eqnarray} and we always assume an SDC/GEM trigger such that $|q^2|=|(p_1-
q_1)^2|\gg\Lambda^2_\rQCD$ and $|q'^2|=|(p_2-q_2)^2|\gg\Lambda^2_\rQCD$.\par
The important fact is that graphs (v)-(vii) in Fig.~(1b) and (v) in
Fig.~(1c) do not
contribute to the infrared singularities in the respective cross sections. More
important is the fact that \be{\rm SUM_{IR}}(\rQCD)=2\alpha_s{\rm Re}B_{QCD}+2
\alpha_s\tilde B_{QCD}(\Kmax)\label{eight}\ee
is also infrared finite: for $m_q=m_{q'
}=m$, for example, we get \be{\rm
SUM_{IR}}(\rQCD)={\alpha_s\over\pi}\sum_{A=\{s
,t,u,s',t',u'\}}(-1)^{\rho(A)}\left(C_FB_{tot}(A)+\Delta C_AB'_{tot}(A)\right)
\label{nine}\ee where \begin{eqnarray*}B_{tot}(A)&=&\log(2\Kmax/\sqrt{|A|})^2(
\ln(|A|/m^2)-1)+\half\ln(|A|/m^2)-1\\ &&\quad-\pi^2/6+\theta(A)\pi^2/2\quad,
\end{eqnarray*} \be B'_{tot}(A)=\log(2\Kmax/\sqrt{|A|})^2\ln(|A|/m^2)
+\half\ln(|A|/m^2)-\pi^2/6+
\theta(A)\pi^2/2\label{ten}\ee and \be\rho(A)=\left\{\begin{array}{lrl}0,&A=s
,s',&t,t'\\ 1,&A=u,u'&\end{array}\right..\label{eleven}\ee (The general
expression for (\ref{nine}) with $m_q\ne m_{q'}$ can be inferred from
Refs.~\cite{ref2}, \cite{ref3} and \cite{ref8} and it is in agreement with the
infrared cancellations in (\ref{nine}).) Thus, for soft gluons with wavelengths
$\gg1/\Lambda_\rQCD$, we find that we may carry through the methods of YFS to
QCD~\cite{ref5}.\par
Specifically, our exponentiated multiple gluon cross section takes the form
\[d\sigma_{\rm exp}=
         \exp[{\rm SUM_{IR}(QCD)}]\sum_{n=0}^\infty\int\prod_{j=1}^n{d^3
k_j\over k_j}\int{d^4y\over(2\pi)^4}e^{iy\cdot(p_1+p_2-q_1-q_2-\sum k_j)+
D_\rQCD}\] \be*\bar\beta_n(k_1,\ldots,k_n){d^3q_1\over q_1^{\,0}}{d^3q_2\over
q_2^{\,0}}\label{twelve}\ee and \be D_\rQCD=\int{d^3k\over k}\tilde S_\rQCD(k)
\left[e^{-iy\cdot k}-\theta(\Kmax-k)\right],\label{thirteen}\ee \[\bar\beta_0=d
\sigma^{\rm(1-loop)}-2\alpha_s{\rm Re}B_\rQCD d\sigma_B,\] \be\bar\beta_1=d
\sigma^{B1}-\tilde S_\rQCD(k)d\sigma_B,\quad\ldots\label{fourteen}\ee where $d
\sigma^{\rm(1-loop)}$ is the $O(\alpha_s)$ cross section for Figs.~(1a) and
(1b)~\cite{ref6}, $d\sigma^{B1}$ is the cross section for
Fig.~(1c)~\cite{ref7}, and $d\sigma_B$ is
the Born cross section for Fig.~(1a). Formulas for the remaining $\bar\beta_n$
can be inferred from Ref.~\cite{ref2}- \cite{ref4} and \cite{ref8}; we show
only
the $\bar\beta_i$ relevant to exact $O(\alpha_s)$ exponentiation for the sake
of
pedagogy. Note that the dummy parameter $\Kmax$ may correspond to an
experimental detector resolution for soft gluon jet energies; we emphasize that
(\ref{twelve}) is independent of $\Kmax$, in complete analogy with the
corresponding circumstance in QED.\par
The hard gluon residuals $\bar\beta_n$ in (\ref{twelve}) can be improved via
the
usual \rngp methods in complete analogy with the \rngp improvement of the QED
YFS hard photon residuals $\bar\beta_n$ in Ref.~\cite{ref8}. This follows from
the fact that QCD, like QED, is a renormalizable quantum field theory, which is
perturbatively calculable so long as the relevant momentum transfer squared is
large compared to $\Lambda^2_\rQCD$. This latter requirement is satisfied for
(\ref{twelve}), where we have in mind the SDC/GEM acceptances at the SSC for
\sqstev\ $pp$ collisions. Following the arguments in Ref.~\cite{ref8}, we
conclude that we get the \rngp improved version of (\ref{twelve}) by making the
substitutions in (\ref{twelve}) (here we use the standard notation~\cite{ref9}
for the running masses $m_i(\lambda)$ and the scaled external momenta
$\{\lambda
\bar p_i\}$) \be\bar\beta_n(\lambda\{\bar p_i\};m_i,\alpha_s)\longrightarrow
\lambda^{2D_\Gamma}\bar\beta_n(\{\bar p_i\};m_i(\lambda),\alpha_s(\lambda))\exp
\left[-\int_1^\lambda d\lambda'2\gamma_\Gamma(\lambda')/\lambda'\right]
\label{fifteen}\ee where $D_\Gamma$ is the respective engineering dimension of
the amplitude from which $\bar\beta_n$ is constructed and $\gamma_\Gamma$
is the anomalous dimension of that amplitude~\cite{ref9}. (We allow that a
finite \rngp transformation may have been used to implement Weinberg's \rngp
operator at a convenient off-shell point and to return our amplitude to the
mass
shell thereafter.) In this way, we arrive at the \rngp improved exponentiated
multiple gluon theory which is entirely analogous to our \rngp improved YFS
theory in Ref.~\cite{ref8} for QED.\par
The results (\ref{twelve}) and (\ref{fifteen}) lend themselves to the same kind
of Monte Carlo event generator realization as did the theory in
Ref.~\cite{ref8}. We have constructed two such event generators from
(\ref{twelve}) and (\ref{fifteen}), one of which, SSCBHLG, treats multiple
gluon
radiation from both initial and final states, and the other of which, SSCYFSG,
treats multiple gluon radiation from the initial state only. These two Monte
Carlo event generators will be discussed in detail elsewhere~\cite{ref4}.\par
In conclusion, in this Letter, we have shown that the infrared singularities in
QCD allow, for our typical SSC process, the same kind of exponentiation as one
can do in QED. This opens the way for the first time for the simulation, via
Monte Carlo event generators, of quantum amplitude based multiple gluon
radiation on an event-by-event basis to all orders in $\alpha_s$. Such results
in the context of SSC/LHC physics will appear elsewhere~\cite{ref4}.\par
$\;$\newline{\bf Acknowledgements:}\par
The authors thank Profs.~F. Gilman and W. Bardeen for the kind hospitality of
the SSC Theory Department, wherein a part of this work was completed.\par
\newpage
{\phantom j}\vfill\begin{center}Figure 1: The process $q+\bbar q{}'\to q+\bbar
q
{}'+(G)$ to $O(\alpha_s)$: (a) Born approximation; (b) $O(\alpha_s)$ virtual
correction; (c) $O(\alpha_s)$ bremsstrahlung process.\end{center}\newpage\pagestyle{empty}\end{document}